\magnification=\magstephalf 

\newbox\SlashedBox 
\def\slashed#1{\setbox\SlashedBox=\hbox{#1}
\hbox to 0pt{\hbox to 1\wd\SlashedBox{\hfil/\hfil}\hss}{#1}}
\def\hboxtosizeof#1#2{\setbox\SlashedBox=\hbox{#1}
\hbox to 1\wd\SlashedBox{#2}}

\def\mathslashed#1{\setbox\SlashedBox=\hbox{$#1$}
\hbox to 0pt{\hbox to 1\wd\SlashedBox{\hfil/\hfil}\hss}#1}

\def\ifsmall{\iffalse}  
\def\titlepagefont{}  

\def\DefineTeXgraphics{%
\special{ps::[global] /TeXgraphics { } def}}  

\def\today{\ifcase\month\or January\or February\or March\or April\or May
\or June\or July\or August\or September\or October\or November\or
December\fi\space\number\day, \number\year}
\def\eatPrefix19{}
\def\Year{\expandafter\eatPrefix\the\year}
\newcount\hours \newcount\minutes
\def\monthname{\ifcase\month\or
January\or February\or March\or April\or May\or June\or July\or
August\or September\or October\or November\or December\fi}
\def\shortmonthname{\ifcase\month\or
Jan\or Feb\or Mar\or Apr\or May\or Jun\or Jul\or
Aug\or Sep\or Oct\or Nov\or Dec\fi}

\def\TimeStamp{\hours\the\time\divide\hours by60%
\minutes -\the\time\divide\minutes by60\multiply\minutes by60%
\advance\minutes by\the\time%
${\rm \shortmonthname}\cdot\if\day<10{}0\fi\the\day\cdot\the\year%
\qquad\the\hours:\if\minutes<10{}0\fi\the\minutes$}




\def\Title#1{%
\vskip 1in{\titlefont\centerline{#1}}\vskip .5in}
 


\newif\ifdraftmode
\newif\ifleftlabels  

\def\nolabels{\def\wrlabeL##1{}\def\eqlabeL##1{}\def\reflabeL##1{}}
\def\writelabels{\def\wrlabeL##1{\leavevmode\vadjust{\rlap{\smash%
{\line{{\escapechar=` \hfill\rlap{\sevenrm\hskip.03in\string##1}}}}}}}%
\def\eqlabeL##1{{\escapechar-1\rlap{\sevenrm\hskip.05in\string##1}}}%
\def\reflabeL##1{\noexpand\rlap{\noexpand\sevenrm[\string##1]}}}
\def\writeleftlabels{\def\wrlabeL##1{\leavevmode\vadjust{\rlap{\smash%
{\line{{\escapechar=` \hfill\rlap{\sevenrm\hskip.03in\string##1}}}}}}}%
\def\eqlabeL##1{{\escapechar-1%
\rlap{\sixrm\hskip.05in\string##1}%
\llap{\sevenrm\string##1\hskip.03in\hbox to \hsize{}}}}%
\def\reflabeL##1{\noexpand\rlap{\noexpand\sevenrm[\string##1]}}}
\nolabels

\newdimen\fullhsize
\newdimen\hstitle
\hstitle=\hsize 
\newdimen\hsbody
\hsbody=\hsize 
\newdimen\hbodyoffset
\hbodyoffset=\hoffset 
\newbox\leftpage
\def\abstract#1{#1}
\def\rotated{\special{ps: landscape}
\magnification=1000  
\baselineskip=14pt
\global\hstitle=9truein\global\hsbody=4.75truein
\global\vsize=7truein\global\voffset=-.31truein
\global\hoffset=-0.54in\global\hbodyoffset=-.54truein
\global\fullhsize=10truein
\def\DefineTeXgraphics{%
\special{ps::[global] 
/TeXgraphics {currentpoint translate 0.7 0.7 scale
              -80 0.72 mul -1000 0.72 mul translate} def}}
\let\lr=L
\def\ifsmall{\iftrue}
\def\titlepagefont{\twelvepoint}
\trueseventeenpoint
\def\almostshipout##1{\if L\lr \count1=1
      \global\setbox\leftpage=##1 \global\let\lr=R
   \else \count1=2
      \shipout\vbox{\hbox to\fullhsize{\box\leftpage\hfil##1}}
      \global\let\lr=L\fi}

\output={\ifnum\count0=1 
 \shipout\vbox{\hbox to \fullhsize{\hfill\pagebody\hfill}}\advancepageno
 \else
 \almostshipout{\leftline{\vbox{\pagebody\makefootline}}}\advancepageno 
 \fi}

\def\abstract##1{{\leftskip=1.5in\rightskip=1.5in ##1\par}} }

\def\linemessage#1{\immediate\write16{#1}}

\global\newcount\secno \global\secno=0
\global\newcount\appno \global\appno=0
\global\newcount\meqno \global\meqno=1
\global\newcount\subsecno \global\subsecno=0
\global\newcount\figno \global\figno=0

\newif\ifAnyCounterChanged
\let\terminator=\relax
\def\normalize#1{\ifx#1\terminator\let\next=\relax\else%
\if#1i\aftergroup i\else\if#1v\aftergroup v\else\if#1x\aftergroup x%
\else\if#1l\aftergroup l\else\if#1c\aftergroup c\else%
\if#1m\aftergroup m\else%
\if#1I\aftergroup I\else\if#1V\aftergroup V\else\if#1X\aftergroup X%
\else\if#1L\aftergroup L\else\if#1C\aftergroup C\else%
\if#1M\aftergroup M\else\aftergroup#1\fi\fi\fi\fi\fi\fi\fi\fi\fi\fi\fi\fi%
\let\next=\normalize\fi%
\next}
\def\makeNormal#1#2{\def\doNormalDef{\edef#1}\begingroup%
\aftergroup\doNormalDef\aftergroup{\normalize#2\terminator\aftergroup}%
\endgroup}

\def\warnIfChanged#1#2{%
\ifundef#1
\else\begingroup%
\edef\oldDefinitionOfCounter{#1}\edef\newDefinitionOfCounter{#2}%
\ifx\oldDefinitionOfCounter\newDefinitionOfCounter%
\else%
\linemessage{Warning: definition of \noexpand#1 has changed.}%
\global\AnyCounterChangedtrue\fi\endgroup\fi}

\def\Section#1{\global\advance\secno by1\relax\global\meqno=1%
\global\subsecno=0%
\bigbreak\bigskip
\centerline{\twelvepoint \bf %
\the\secno. #1}%
\par\nobreak\medskip\nobreak}
\def\tagsection#1{%
\warnIfChanged#1{\the\secno}%
\xdef#1{\the\secno}%
\ifWritingAuxFile\immediate\write\auxfile{\noexpand\xdef\noexpand#1{#1}}\fi%
}
\def\section{\Section}
\def\Subsection#1{\global\advance\subsecno by1\relax\medskip %
\leftline{\bf\the\secno.\the\subsecno\ #1}%
\par\nobreak\smallskip\nobreak}
\def\tagsubsection#1{%
\warnIfChanged#1{\the\secno.\the\subsecno}%
\xdef#1{\the\secno.\the\subsecno}%
\ifWritingAuxFile\immediate\write\auxfile{\noexpand\xdef\noexpand#1{#1}}\fi%
}

\def\subsection{\Subsection}

\def\romappno{\uppercase\expandafter{\romannumeral\appno}}
\def\makeNormalizedRomappno{%
\expandafter\makeNormal\expandafter\normalizedromappno%
\expandafter{\romannumeral\appno}%
\edef\normalizedromappno{\uppercase{\normalizedromappno}}}
\def\Appendix#1{\global\advance\appno by1\relax\global\meqno=1\global\secno=0%
\global\subsecno=0%
\bigbreak\bigskip
\centerline{\twelvepoint \bf Appendix %
\romappno. #1}%
\par\nobreak\medskip\nobreak}
\def\tagappendix#1{\makeNormalizedRomappno%
\warnIfChanged#1{\normalizedromappno}%
\xdef#1{\normalizedromappno}%
\ifWritingAuxFile\immediate\write\auxfile{\noexpand\xdef\noexpand#1{#1}}\fi%
}
\def\appendix{\Appendix}
\def\Subappendix#1{\global\advance\subsecno by1\relax\medskip %
\leftline{\bf\romappno.\the\subsecno\ #1}%
\par\nobreak\smallskip\nobreak}
\def\tagsubappendix#1{\makeNormalizedRomappno%
\warnIfChanged#1{\normalizedromappno.\the\subsecno}%
\xdef#1{\normalizedromappno.\the\subsecno}%
\ifWritingAuxFile\immediate\write\auxfile{\noexpand\xdef\noexpand#1{#1}}\fi%
}

\def\eqn#1{\makeNormalizedRomappno%
\ifnum\secno>0%
  \warnIfChanged#1{\the\secno.\the\meqno}%
  \eqno(\the\secno.\the\meqno)\xdef#1{\the\secno.\the\meqno}%
     \global\advance\meqno by1
\else\ifnum\appno>0%
  \warnIfChanged#1{\normalizedromappno.\the\meqno}%
  \eqno({\rm\romappno}.\the\meqno)%
      \xdef#1{\normalizedromappno.\the\meqno}%
     \global\advance\meqno by1
\else%
  \warnIfChanged#1{\the\meqno}%
  \eqno(\the\meqno)\xdef#1{\the\meqno}%
     \global\advance\meqno by1
\fi\fi%
\eqlabeL#1%
\ifWritingAuxFile\immediate\write\auxfile{\noexpand\xdef\noexpand#1{#1}}\fi%
}
\def\defeqn#1{\makeNormalizedRomappno%
\ifnum\secno>0%
  \warnIfChanged#1{\the\secno.\the\meqno}%
  \xdef#1{\the\secno.\the\meqno}%
     \global\advance\meqno by1
\else\ifnum\appno>0%
  \warnIfChanged#1{\normalizedromappno.\the\meqno}%
  \xdef#1{\normalizedromappno.\the\meqno}%
     \global\advance\meqno by1
\else%
  \warnIfChanged#1{\the\meqno}%
  \xdef#1{\the\meqno}%
     \global\advance\meqno by1
\fi\fi%
\eqlabeL#1%
\ifWritingAuxFile\immediate\write\auxfile{\noexpand\xdef\noexpand#1{#1}}\fi%
}
\def\anoneqn{\makeNormalizedRomappno%
\ifnum\secno>0
  \eqno(\the\secno.\the\meqno)%
     \global\advance\meqno by1
\else\ifnum\appno>0
  \eqno({\rm\normalizedromappno}.\the\meqno)%
     \global\advance\meqno by1
\else
  \eqno(\the\meqno)%
     \global\advance\meqno by1
\fi\fi%
}
\def\mfig#1#2{\ifx#20
\else\global\advance\figno by1%
\relax#1\the\figno%
\warnIfChanged#2{\the\figno}%
\xdef#2{\the\figno}%
\reflabeL#2%
\ifWritingAuxFile\immediate\write\auxfile{\noexpand\xdef\noexpand#2{#2}}\fi\fi%
}

\def\fig#1{\mfig{fig.\ ~}#1}

\catcode`@=11 

\newif\ifFiguresInText\FiguresInTexttrue
\newif\if@FigureFileCreated
\newwrite\capfile
\newwrite\figfile

\newif\ifcaption
\captiontrue
\def\captionsize{\tenrm}
\def\PlaceTextFigure#1#2#3#4{%
\vskip 0.5truein%
#3\hfil\epsfbox{#4}\hfil\break%
\ifcaption\hfil\vbox{\captionsize Figure #1. #2}\hfil\fi%
\vskip10pt}
\def\PlaceEndFigure#1#2{%
\epsfxsize=\hsize\epsfbox{#2}\vfill\centerline{Figure #1.}\eject}

\def\LoadFigure#1#2#3#4{%
\ifundef#1{\phantom{\mfig{}#1}}\else
\fi%
\ifFiguresInText
\PlaceTextFigure{#1}{#2}{#3}{#4}%
\else
\if@FigureFileCreated\else%
\immediate\openout\capfile=\jobname.caps%
\immediate\openout\figfile=\jobname.figs%
@FigureFileCreatedtrue\fi%
\immediate\write\capfile{\noexpand\item{Figure \noexpand#1.\ }{#2}\vskip10pt}%
\immediate\write\figfile{\noexpand\PlaceEndFigure\noexpand#1{\noexpand#4}}%
\fi}

\def\listfigs{\ifFiguresInText\else%
\vfill\eject\immediate\closeout\capfile
\immediate\closeout\figfile%
\centerline{{\bf Figures}}\bigskip\frenchspacing%
\catcode`@=11 
\def\captionsize{\tenrm}
\input \jobname.caps\vfill\eject\nonfrenchspacing%
\catcode`\@=\active
\catcode`@=12  
\input\jobname.figs\fi}

\font\ninerm=cmr9
\font\eightrm=cmr8
\font\sixrm=cmr6

\def\loadtrueseventeenpoint{
 \font\seventeenrm=cmr10 at 17.28truept
 \font\seventeeni=cmmi10 at 17.28truept
 \font\seventeenbf=cmbx10 at 17.28truept
 \font\seventeenit=cmti10 at 17.28truept
 \font\seventeensl=cmsl10 at 17.28truept
 \font\seventeensy=cmsy10 at 17.28truept
}
\def\loadfourteenpoint{
\font\fourteenrm=cmr10 at 14.4pt
\font\fourteeni=cmmi10 at 14.4pt
\font\fourteenit=cmti10 at 14.4pt
\font\fourteensl=cmsl10 at 14.4pt
\font\fourteensy=cmsy10 at 14.4pt
\font\fourteenbf=cmbx10 at 14.4pt
}
\def\loadtruetwelvepoint{
\font\twelverm=cmr10 at 12truept
\font\twelvei=cmmi10 at 12truept
\font\twelveit=cmti10 at 12truept
\font\twelvesl=cmsl10 at 12truept
\font\twelvesy=cmsy10 at 12truept
\font\twelvebf=cmbx10 at 12truept
}

\font\ninei=cmmi9
\font\eighti=cmmi8
\font\sixi=cmmi6
\skewchar\ninei='177 \skewchar\eighti='177 \skewchar\sixi='177

\font\ninesy=cmsy9
\font\eightsy=cmsy8
\font\sixsy=cmsy6
\skewchar\ninesy='60 \skewchar\eightsy='60 \skewchar\sixsy='60

\font\ninebf=cmbx9
\font\eightbf=cmbx8
\font\sixbf=cmbx6

\font\ninett=cmtt9
\font\eighttt=cmtt8

\hyphenchar\tentt=-1 
\hyphenchar\ninett=-1
\hyphenchar\eighttt=-1         

\font\ninesl=cmsl9
\font\eightsl=cmsl8

\font\nineit=cmti9
\font\eightit=cmti8

                      
\newskip\ttglue
\def\tenpoint{\def\rm{\fam0\tenrm}%
  \textfont0=\tenrm \scriptfont0=\sevenrm \scriptscriptfont0=\fiverm
  \textfont1=\teni \scriptfont1=\seveni \scriptscriptfont1=\fivei
  \textfont2=\tensy \scriptfont2=\sevensy \scriptscriptfont2=\fivesy
  \textfont3=\tenex \scriptfont3=\tenex \scriptscriptfont3=\tenex
  \def\it{\fam\itfam\tenit}\textfont\itfam=\tenit
  \def\sl{\fam\slfam\tensl}\textfont\slfam=\tensl
  \def\bf{\fam\bffam\tenbf}\textfont\bffam=\tenbf \scriptfont\bffam=\sevenbf
  \scriptscriptfont\bffam=\fivebf
  \normalbaselineskip=12pt
  \let\sc=\eightrm
  \let\big=\tenbig
  \setbox\strutbox=\hbox{\vrule height8.5pt depth3.5pt width\z@}%
  \normalbaselines\rm}

\def\twelvepoint{\def\rm{\fam0\twelverm}%
  \textfont0=\twelverm \scriptfont0=\ninerm \scriptscriptfont0=\sevenrm
  \textfont1=\twelvei \scriptfont1=\ninei \scriptscriptfont1=\seveni
  \textfont2=\twelvesy \scriptfont2=\ninesy \scriptscriptfont2=\sevensy
  \textfont3=\tenex \scriptfont3=\tenex \scriptscriptfont3=\tenex
  \def\it{\fam\itfam\twelveit}\textfont\itfam=\twelveit
  \def\sl{\fam\slfam\twelvesl}\textfont\slfam=\twelvesl
  \def\bf{\fam\bffam\twelvebf}\textfont\bffam=\twelvebf%
  \scriptfont\bffam=\ninebf
  \scriptscriptfont\bffam=\sevenbf
  \normalbaselineskip=12pt
  \let\sc=\eightrm
  \let\big=\tenbig
  \setbox\strutbox=\hbox{\vrule height8.5pt depth3.5pt width\z@}%
  \normalbaselines\rm}

\def\fourteenpoint{\def\rm{\fam0\fourteenrm}%
  \textfont0=\fourteenrm \scriptfont0=\tenrm \scriptscriptfont0=\sevenrm
  \textfont1=\fourteeni \scriptfont1=\teni \scriptscriptfont1=\seveni
  \textfont2=\fourteensy \scriptfont2=\tensy \scriptscriptfont2=\sevensy
  \textfont3=\tenex \scriptfont3=\tenex \scriptscriptfont3=\tenex
  \def\it{\fam\itfam\fourteenit}\textfont\itfam=\fourteenit
  \def\sl{\fam\slfam\fourteensl}\textfont\slfam=\fourteensl
  \def\bf{\fam\bffam\fourteenbf}\textfont\bffam=\fourteenbf%
  \scriptfont\bffam=\tenbf
  \scriptscriptfont\bffam=\sevenbf
  \normalbaselineskip=17pt
  \let\sc=\elevenrm
  \let\big=\tenbig                                          
  \setbox\strutbox=\hbox{\vrule height8.5pt depth3.5pt width\z@}%
  \normalbaselines\rm}

\def\seventeenpoint{\def\rm{\fam0\seventeenrm}%
  \textfont0=\seventeenrm \scriptfont0=\fourteenrm \scriptscriptfont0=\tenrm
  \textfont1=\seventeeni \scriptfont1=\fourteeni \scriptscriptfont1=\teni
  \textfont2=\seventeensy \scriptfont2=\fourteensy \scriptscriptfont2=\tensy
  \textfont3=\tenex \scriptfont3=\tenex \scriptscriptfont3=\tenex
  \def\it{\fam\itfam\seventeenit}\textfont\itfam=\seventeenit
  \def\sl{\fam\slfam\seventeensl}\textfont\slfam=\seventeensl
  \def\bf{\fam\bffam\seventeenbf}\textfont\bffam=\seventeenbf%
  \scriptfont\bffam=\fourteenbf
  \scriptscriptfont\bffam=\twelvebf
  \normalbaselineskip=21pt
  \let\sc=\fourteenrm
  \let\big=\tenbig                                          
  \setbox\strutbox=\hbox{\vrule height 12pt depth 6pt width\z@}%
  \normalbaselines\rm}

\def\ninepoint{\def\rm{\fam0\ninerm}%
  \textfont0=\ninerm \scriptfont0=\sixrm \scriptscriptfont0=\fiverm
  \textfont1=\ninei \scriptfont1=\sixi \scriptscriptfont1=\fivei
  \textfont2=\ninesy \scriptfont2=\sixsy \scriptscriptfont2=\fivesy
  \textfont3=\tenex \scriptfont3=\tenex \scriptscriptfont3=\tenex
  \def\it{\fam\itfam\nineit}\textfont\itfam=\nineit
  \def\sl{\fam\slfam\ninesl}\textfont\slfam=\ninesl
  \def\bf{\fam\bffam\ninebf}\textfont\bffam=\ninebf \scriptfont\bffam=\sixbf
  \scriptscriptfont\bffam=\fivebf
  \normalbaselineskip=11pt
  \let\sc=\sevenrm
  \let\big=\ninebig
  \setbox\strutbox=\hbox{\vrule height8pt depth3pt width\z@}%
  \normalbaselines\rm}

\def\eightpoint{\def\rm{\fam0\eightrm}%
  \textfont0=\eightrm \scriptfont0=\sixrm \scriptscriptfont0=\fiverm%
  \textfont1=\eighti \scriptfont1=\sixi \scriptscriptfont1=\fivei%
  \textfont2=\eightsy \scriptfont2=\sixsy \scriptscriptfont2=\fivesy%
  \textfont3=\tenex \scriptfont3=\tenex \scriptscriptfont3=\tenex%
  \def\it{\fam\itfam\eightit}\textfont\itfam=\eightit%
  \def\sl{\fam\slfam\eightsl}\textfont\slfam=\eightsl%
  \def\bf{\fam\bffam\eightbf}\textfont\bffam=\eightbf \scriptfont\bffam=\sixbf%
  \scriptscriptfont\bffam=\fivebf%
  \normalbaselineskip=9pt%
  \let\sc=\sixrm%
  \let\big=\eightbig%
  \setbox\strutbox=\hbox{\vrule height7pt depth2pt width\z@}%
  \normalbaselines\rm}

\def\tenbig#1{{\hbox{$\left#1\vbox to8.5pt{}\right.\n@space$}}}
\def\ninebig#1{{\hbox{$\textfont0=\tenrm\textfont2=\tensy
  \left#1\vbox to7.25pt{}\right.\n@space$}}}
\def\eightbig#1{{\hbox{$\textfont0=\ninerm\textfont2=\ninesy
  \left#1\vbox to6.5pt{}\right.\n@space$}}}

\def\footnote#1{\edef\@sf{\spacefactor\the\spacefactor}#1\@sf
      \insert\footins\bgroup\eightpoint
      \interlinepenalty100 \let\par=\endgraf
        \leftskip=\z@skip \rightskip=\z@skip
        \splittopskip=10pt plus 1pt minus 1pt \floatingpenalty=20000
        \smallskip\item{#1}\bgroup\strut\aftergroup\@foot\let\next}
\skip\footins=12pt plus 2pt minus 4pt 
\dimen\footins=30pc 

\newinsert\margin
\dimen\margin=\maxdimen
\def\titlefont{\seventeenpoint}
\loadtruetwelvepoint 
\loadtrueseventeenpoint

\def\eatOne#1{}
\def\ifundef#1{\expandafter\ifx%
\csname\expandafter\eatOne\string#1\endcsname\relax}
\def\notTrue{\iffalse}\def\isTrue{\iftrue}
\def\ifdef#1{{\ifundef#1%
\aftergroup\notTrue\else\aftergroup\isTrue\fi}}
\def\use#1{\ifundef#1\linemessage{Warning: \string#1 is undefined.}%
{\tt \string#1}\else#1\fi}



%
\catcode`"=11
\let\quote="
\catcode`"=12
\chardef\foo="22
\global\newcount\refno \global\refno=1
\newwrite\rfile
\newlinechar=`\^^J
\def\@ref#1#2{\the\refno\n@ref#1{#2}}
\def\h@ref#1#2#3{\href{#3}{\the\refno}\n@ref#1{#2}}
\def\n@ref#1#2{\xdef#1{\the\refno}%
\ifnum\refno=1\immediate\openout\rfile=\jobname.refs\fi%
\immediate\write\rfile{\noexpand\item{[\noexpand#1]\ }#2.}%
\global\advance\refno by1}
\def\nref{\n@ref} 
\def\ref{\@ref}   
\def\hrref{\h@ref}
\def\lref#1#2{\the\refno\xdef#1{\the\refno}%
\ifnum\refno=1\immediate\openout\rfile=\jobname.refs\fi%
\immediate\write\rfile{\noexpand\item{[\noexpand#1]\ }#2\semi}%
\global\advance\refno by1}
\def\cref#1{\immediate\write\rfile{#1\semi}}

\def\preref#1#2{\gdef#1{\@ref#1{#2}}}

\def\semi{;\hfil\noexpand\break}

\def\listrefs{\vfill\eject\immediate\closeout\rfile
\centerline{{\bf References}}\bigskip\frenchspacing%
\input \jobname.refs\vfill\eject\nonfrenchspacing}

\def\inputAuxIfPresent#1{\immediate\openin1=#1
\ifeof1\message{No file \auxfileName; I'll create one.
}\else\closein1\relax\input\auxfileName\fi%
}
\def\NPB{Nucl.\ Phys.\ B}
\def\PRL{Phys.\ Rev.\ Lett.\ }
\def\PRD{Phys.\ Rev.\ D}
\def\PLB{Phys.\ Lett.\ B}




\newif\ifWritingAuxFile
\newwrite\auxfile
\def\SetUpAuxFile{%
\xdef\auxfileName{\jobname.aux}%
\inputAuxIfPresent{\auxfileName}%
\WritingAuxFiletrue%
\immediate\openout\auxfile=\auxfileName}


\def\LB{\left[}\def\RB{\right]}


\catcode`\@=\active
\catcode`@=12  
\catcode`\"=\active


\def\Tr{\mathop{\rm Tr}\nolimits}

\def\pol{\varepsilon}

\def\ksl{\slashed{k}}

\def\spa#1.#2{\left\langle#1\,#2\right\rangle}
\def\spb#1.#2{\left[#1\,#2\right]}
\def\lor#1.#2{\left(#1\,#2\right)}
\def\sand#1.#2.#3{%
\left\langle\smash{#1}{\vphantom1}^{-}\right|{#2}%
\left|\smash{#3}{\vphantom1}^{-}\right\rangle}
\def\sandp#1.#2.#3{%
\left\langle\smash{#1}{\vphantom1}^{-}\right|{#2}%
\left|\smash{#3}{\vphantom1}^{+}\right\rangle}
\def\sandpp#1.#2.#3{%
\left\langle\smash{#1}{\vphantom1}^{+}\right|{#2}%
\left|\smash{#3}{\vphantom1}^{+}\right\rangle}
\def\sandpm#1.#2.#3{%
\left\langle\smash{#1}{\vphantom1}^{+}\right|{#2}%
\left|\smash{#3}{\vphantom1}^{-}\right\rangle}
\def\sandmp#1.#2.#3{%
\left\langle\smash{#1}{\vphantom1}^{-}\right|{#2}%
\left|\smash{#3}{\vphantom1}^{+}\right\rangle}
\catcode`@=11  
\def\meqalign#1{\,\vcenter{\openup1\jot\m@th
   \ialign{\strut\hfil$\displaystyle{##}$ && $\displaystyle{{}##}$\hfil
             \crcr#1\crcr}}\,}
\catcode`@=12  


\def\Tr{\mathop{\rm Tr}\nolimits}

\def\eps{\epsilon}

\def\pol{\varepsilon}

\def\dl^#1_#2{\delta^{#1}{}_{#2}}

\def\Ord{{\cal O}}
\def\ord{{\eightpoint\cal O}}

\catcode`@=11  
\def\meqalign#1{\,\vcenter{\openup1\jot\m@th
   \ialign{\strut\hfil$\displaystyle{##}$ && $\displaystyle{{}##}$\hfil
             \crcr#1\crcr}}\,}
\catcode`@=12  


\baselineskip 15pt
\overfullrule 0.5pt

\newread\epsffilein    
\newif\ifepsffileok    
\newif\ifepsfbbfound   
\newif\ifepsfverbose   
\newdimen\epsfxsize    
\newdimen\epsfysize    
\newdimen\epsftsize    
\newdimen\epsfrsize    
\newdimen\epsftmp      
\newdimen\pspoints     
\pspoints=1bp          
\epsfxsize=0pt         
\epsfysize=0pt         
\def\epsfbox#1{\global\def\epsfllx{72}\global\def\epsflly{72}%
   \global\def\epsfurx{540}\global\def\epsfury{720}%
   \def\lbracket{[}\def\testit{#1}\ifx\testit\lbracket
   \let\next=\epsfgetlitbb\else\let\next=\epsfnormal\fi\next{#1}}%
\def\epsfgetlitbb#1#2 #3 #4 #5]#6{\epsfgrab #2 #3 #4 #5 .\\%
   \epsfsetgraph{#6}}%
\def\epsfnormal#1{\epsfgetbb{#1}\epsfsetgraph{#1}}%
\def\epsfgetbb#1{%
%
%
\openin\epsffilein=#1
\ifeof\epsffilein\errmessage{I couldn't open #1, will ignore it}\else
%
%
   {\epsffileoktrue \chardef\other=12
    \def\do##1{\catcode`##1=\other}\dospecials \catcode`\ =10
    \loop
       \read\epsffilein to \epsffileline
       \ifeof\epsffilein\epsffileokfalse\else
%
%
          \expandafter\epsfaux\epsffileline:. \\%
       \fi
   \ifepsffileok\repeat
   \ifepsfbbfound\else
    \ifepsfverbose\message{No bounding box comment in #1; using defaults}\fi\fi
   }\closein\epsffilein\fi}%
%
%
\def\epsfclipstring{}
\def\epsfsetgraph#1{%
   \epsfrsize=\epsfury\pspoints
   \advance\epsfrsize by-\epsflly\pspoints
   \epsftsize=\epsfurx\pspoints
   \advance\epsftsize by-\epsfllx\pspoints
%
%
   \epsfxsize\epsfsize\epsftsize\epsfrsize
   \ifnum\epsfxsize=0 \ifnum\epsfysize=0
      \epsfxsize=\epsftsize \epsfysize=\epsfrsize
      \epsfrsize=0pt
%
%
     \else\epsftmp=\epsftsize \divide\epsftmp\epsfrsize
       \epsfxsize=\epsfysize \multiply\epsfxsize\epsftmp
       \multiply\epsftmp\epsfrsize \advance\epsftsize-\epsftmp
       \epsftmp=\epsfysize
       \loop \advance\epsftsize\epsftsize \divide\epsftmp 2
       \ifnum\epsftmp>0
          \ifnum\epsftsize<\epsfrsize\else
             \advance\epsftsize-\epsfrsize \advance\epsfxsize\epsftmp \fi
       \repeat
       \epsfrsize=0pt
     \fi
   \else \ifnum\epsfysize=0
     \epsftmp=\epsfrsize \divide\epsftmp\epsftsize
     \epsfysize=\epsfxsize \multiply\epsfysize\epsftmp   
     \multiply\epsftmp\epsftsize \advance\epsfrsize-\epsftmp
     \epsftmp=\epsfxsize
     \loop \advance\epsfrsize\epsfrsize \divide\epsftmp 2
     \ifnum\epsftmp>0
        \ifnum\epsfrsize<\epsftsize\else
           \advance\epsfrsize-\epsftsize \advance\epsfysize\epsftmp \fi
     \repeat
     \epsfrsize=0pt
    \else
     \epsfrsize=\epsfysize
    \fi
   \fi
%
%
   \ifepsfverbose\message{#1: width=\the\epsfxsize, height=\the\epsfysize}\fi
   \epsftmp=10\epsfxsize \divide\epsftmp\pspoints
   \vbox to\epsfysize{\vfil\hbox to\epsfxsize{%
      \ifnum\epsfrsize=0\relax
        \includegraphics{#1}%
      \else
        \epsfrsize=10\epsfysize \divide\epsfrsize\pspoints
        \includegraphics{#1}%
      \fi
      \hfil}}%
\global\epsfxsize=0pt\global\epsfysize=0pt}%
%
%
{\catcode`\%=12 \global\let\epsfpercent=
%
%
\long\def\epsfaux#1#2:#3\\{\ifx#1\epsfpercent
   \def\testit{#2}\ifx\testit\epsfbblit
      \epsfgrab #3 . . . \\%
      \epsffileokfalse
      \global\epsfbbfoundtrue
   \fi\else\ifx#1\par\else\epsffileokfalse\fi\fi}%
%
%
\def\epsfempty{}%
\def\epsfgrab #1 #2 #3 #4 #5\\{%
\global\def\epsfllx{#1}\ifx\epsfllx\epsfempty
      \epsfgrab #2 #3 #4 #5 .\\\else
   \global\def\epsflly{#2}%
   \global\def\epsfurx{#3}\global\def\epsfury{#4}\fi}%
%
%
\def\epsfsize#1#2{\epsfxsize}
%
%


\hfuzz 30 pt

\def\tree{{\rm tree}}
\def\dlips{{\rm dLIPS}}

\def\Tr{{\rm Tr}}
\def\tr{{\rm tr}}

\def\Psl{\slashed{P}}
\def\ksl{\slashed{k}}
\def\ksl{\slashed{k}}
\def\lsl{\mathslashed{\ell}}

\def\as#1{a_{\sigma(#1)}}
\def\sig#1{\sigma(#1)}
\def\hf{{\textstyle 1\over2}}
\def\to{\rightarrow}
\def\e{\eps}

\def\sand#1.#2.#3{%
  \left\langle\smash{#1}{\vphantom1}\right|{#2}%
  \left|\smash{#3}{\vphantom1}\right\rangle}
\def\sandp#1.#2.#3{%
  \left\langle\smash{#1}{\vphantom1}^{-}\right|{#2}%
  \left|\smash{#3}{\vphantom1}^{+}\right\rangle}
\def\sandpp#1.#2.#3{%
  \left\langle\smash{#1}{\vphantom1}^{+}\right|{#2}%
  \left|\smash{#3}{\vphantom1}^{+}\right\rangle}
\def\sandmm#1.#2.#3{%
  \left\langle\smash{#1}{\vphantom1}^{-}\right|{#2}%
  \left|\smash{#3}{\vphantom1}^{-}\right\rangle}
\def\sandpm#1.#2.#3{%
  \left\langle\smash{#1}{\vphantom1}^{+}\right|{#2}%
  \left|\smash{#3}{\vphantom1}^{-}\right\rangle}
\def\sandmp#1.#2.#3{%
  \left\langle\smash{#1}{\vphantom1}^{-}\right|{#2}%
  \left|\smash{#3}{\vphantom1}^{+}\right\rangle}

\SetUpAuxFile
 
\loadfourteenpoint
\hfuzz 40 pt


\preref\StringBased{
Z. Bern and D.A.\ Kosower, Phys.\ Rev.\ Lett.\ 66:1669 (1991)\semi
Z. Bern, Phys.\ Lett.\ 296B:85 (1992)\semi
Z. Bern and D.C.\ Dunbar,  Nucl.\ Phys.\ B379:562 (1992)}

\preref\Long{Z. Bern and D.A.\ Kosower, Nucl.\ Phys.\ B379:451 (1992)}

\preref\BKLoopColor{Z. Bern and D.A.\ Kosower, Nucl.\ Phys.\ B362:389 (1991)}
 
\preref\TreeColor{J.E.\ Paton and H.M.\ Chan, Nucl.\ Phys.\ B10:516
(1969)\semi
F.A.\ Berends and W.T.\ Giele, \NPB 294:700 (1987)\semi
M.\ Mangano, \NPB 309:461 (1988)}

\preref\Fermion{Z. Bern, L. Dixon and D.A.\ Kosower, \NPB 437:259
(1995), hep-ph/9409393}

\preref\Gravity{Z. Bern, D.C. Dunbar and T. Shimada,
Phys.\ Lett.\ 312B:277 (1993), hep-th/9307001\semi
D.C.\  Dunbar and P.S.\ Norridge,
Nucl.\ Phys.\ B433:181 (1995), hep-th/9408014; hep-th/9512084}

\preref\FiveGluon{Z. Bern, L. Dixon and D.A.\ Kosower, Phys.\ Rev.\
Lett.\
70:2677 (1993), hep-ph/9302280}

\preref\GSB{M.B.\ Green, J.H.\ Schwarz and L.\ Brink,
 Nucl.\ Phys.\ B198:474 (1982)}

\preref\GSW{M.B.\ Green, J.H.\ Schwarz,
and E.\ Witten, {\it Superstring Theory} (Cambridge University
Press) (1987)}

\preref\BerendsGravity{F.A. Berends, W.T.\ Giele and H. Kuijf,
Phys. Lett.\ 211B:91 (1988)}

\preref\Cutkosky{L.D.\ Landau, Nucl.\ Phys.\ 13:181 (1959)\semi
S. Mandelstam, Phys.\ Rev.\ 112:1344 (1958), 115:1741 (1959)\semi
R.E.\ Cutkosky, J.\ Math.\ Phys.\ 1:429 (1960)\semi
V. Constanti, B. DeTollis and G. Pistoni, Nuovo Cimento
2A:733 (1971)\semi
 B. DeTollis, M. Lusignoli and G. Pistoni, Nuovo Cimento
 32A:227 (1976)}
      
\preref\SusyFour{Z. Bern, L. Dixon, D.C. Dunbar and D.A. Kosower,
Nucl.\ Phys.\ B425:217 (1994), hep-ph/9403226}
	
\preref\SusyOne{Z. Bern, L. Dixon, D.C. Dunbar and D.A. Kosower,
Nucl.\ Phys.\ B435:59 (1995), hep-ph/9409265}

\preref\Massive{
Z.\ Bern and A.G.\ Morgan, Nucl.\ Phys.\ B467:479 (1996),
hep-ph/9511336}
	 
\preref\KST{Z. Kunszt, A. Signer and Z. Tr\'ocs\'anyi,
Nucl.\ Phys.\ B420:550 (1994), hep-ph/9401294}
	   
\preref\Cangemi{
D. Cangemi, preprint hep-th/9605208}

\preref\CangemiConf{
D. Cangemi,  preprint hep-th/9610021}

\preref\Bardeen{
W.A. Bardeen, preprint FERMILAB-CONF-95-379-T}

\preref\Selivanov{K.G. Selivanov, preprint hep-ph/9604206}

\preref\Nair{V.P. Nair, \PLB 214:215 (1988)}

\preref\DNS{C.N. Yang, Phys. Rev. Lett. 38:1377 (1977)\semi
S. Donaldson, Proc. Lond. Math. Soc. 50:1 (1985)\semi
V.P. Nair and J. Schiff, \PLB 246:423 (1990); Nucl.\ Phys. B371:329 (1992)}

\preref\NTwoString{H. Ooguri and C. Vafa, Nucl.\ Phys.\ B361:469 (1991);
Nucl.\ Phys.\ B367:83 (1991); 
Nucl.\ Phys.\ B451:121 (1995), hep-th/9505183\semi
N. Berkovits and C. Vafa, Nucl.\ Phys.\ B433:123 (1995),
hep-th/9407190} 
 
\preref\Siegel{G. Chalmers and W. Siegel, preprint hep-th/9606061}

\preref\LCG{A.N. Leznov, Theor. Math. Phys. 73:1233 (1988)\semi
A.N. Leznov and M.A. Mukhtarov, J. Math. Phys. 28:2574 (1987)\semi
A. Parkes, \PLB 286:265 (1992)\semi
W. Siegel, \PRD 46:R3235 (1992)}

\preref\AllPlus{Z. Bern, G. Chalmers, L. Dixon and D.A.\ Kosower,
Phys.\ Rev.\ Lett.\ 72:2134 (1994), hep-ph/9312333}

\preref\Schwinger{J. Schwinger, Phys.\ Rev. 82:664 (1951)}  

\preref\FirstQ{
E.S.\ Fradkin and A.A.\ Tseytlin, Phys. Lett. 158B:316
(1985);
163B:123 (1985); Nucl. Phys. B261:1 (1985)\semi
M. Strassler, Nucl.\ Phys.\ {B385}:145 (1992), hep-ph/9205205\semi
M.G. Schmidt and C. Schubert, Phys.\ Lett.\ 318B:438
(1993), hep-th/9309055\semi
D.G.C.\ McKeon, Ann. Phys. (N.Y.) 224:139 (1993)}

\preref\DimReg{G. 't\ Hooft and M. Veltman,
Nucl.\ Phys.\ B44:189 (1972)}
 
\preref\SusyReg{W. Siegel, Phys.\ Lett.\ 84B:193 (1979)\semi
D.M.\ Capper, D.R.T.\ Jones and P. van Nieuwenhuizen,
Nucl.\ Phys.\
B167:479 (1980)\semi
L.V.\ Avdeev and A.A.\ Vladimirov, Nucl.\ Phys.\ B219:262
(1983)\semi
I.\ Jack, D.R.T.\ Jones and K.L. Roberts,  Z. Phys.
C63:151 (1994)}
	       
\preref\Factorization{Z.\ Bern and G.\ Chalmers,
Nucl.\ Phys.\ B447:465 (1995), hep-ph/9503236}

\preref\SchemeConversion{Z. Kunszt, A. Signer and 
Z. Tr\'ocs\'anyi, Nucl.\ Phys.\ B411:397 (1994), hep-ph/9305239\semi
S. Catani, M.H. Seymour and Z. Tr\'ocs\'anyi, preprint hep-ph/9610553}

\preref\SpinorHelicity{
F.A.\ Berends, R.\ Kleiss, P.\ De Causmaecker, R.\ Gastmans and T.T.\ Wu,
Phys.\ Lett.\ 103B:124 (1981)\semi
P. De Causmaeker, R. Gastmans,  W.\ Troost and  T.T.\ Wu,
Nucl. Phys. B206:53 (1982)\semi
R.\ Kleiss and W.\ J.\ Stirling,
Nucl.\ Phys.\ B262:235 (1985)\semi
J.F.\ Gunion and Z.\ Kunszt, Phys.\ Lett.\ 161B:333
(1985)\semi
Z.\ Xu, D.-H.\ Zhang and L. Chang, Nucl.\ Phys.\
B291:392 (1987)}

\preref\Mahlon{G.D.\ Mahlon, Phys.\ Rev. D49 (1994) 2197,
hep-ph/9311213; 
Phys.\ Rev. D49 (1994) 4438, hep-ph/9312276}
		 
\preref\SWI{
M.T.\ Grisaru, H.N.\ Pendleton and P.\ van Nieuwenhuizen,
Phys. Rev. {D15}:996 (1977)\semi
M.T. Grisaru and H.N. Pendleton, Nucl.\ Phys.\ B124:81 (1977)\semi
S.J. Parke and T. Taylor, Phys.\ Lett.\ 157B:81 (1985)}
 
\preref\ManganoReview{M.L.\ Mangano and S.J. Parke, Phys.\ Rep.\
{200}:301 (1991)\semi
L. Dixon, hep-ph/9601359,
in {\it Proceedings of Theoretical Advanced Study Institute in
Elementary Particle Physics (TASI 95)}, ed.\ D.E.\ Soper}

\preref\TasiZvi{Z. Bern, hep-ph/9304249, in {\it Proceedings of Theoretical
Advanced Study Institute in High Energy Physics (TASI 92)},
eds.\ J. Harvey and J. Polchinski (World Scientific, 1993),
hep-ph/9304249}

\preref\WeakInt{Z.\ Bern and A.G.\ Morgan, Phys.\ Rev.\ D49:6155 (1994), 
hep-ph/9312218}

\preref\Integrals{Z. Bern, L. Dixon and D.A.\ Kosower,
\NPB 412:751 (1994), hep-ph/9306240}

\preref\MVNV{
D.B. Melrose, Il Nuovo Cimento 40A:181 (1965)\semi
W. van Neerven and J.A.M. Vermaseren, Phys.\ Lett.\ 137B:241 (1984)}

\preref\ParkeTaylor{S.J.\ Parke and T.R.\ Taylor, 
\NPB 269:410 (1986), \PRL 56:2459
(1986)}

\preref\Recursive{F.A.\ Berends and W.T.\ Giele, Nucl.\ Phys.\ B306:759 
(1988)\semi
D.A.\ Kosower, Nucl.\ Phys.\ B335:23 (1990)\semi
G. Mahlon and T.-M.\ Yan,  Phys.\ Rev.\ D47:1776 (1993),
hep-ph/9210213\semi
G. Mahlon, Phys.\ Rev.\ D47:1812 (1993), hep-ph/9210214\semi
G. Mahlon, T.-M.\ Yan and C. Dunn, Phys.\ Rev.\ D48:1337 (1993),
hep-ph/9210212\semi
C. Kim and  V.P. Nair, preprint hep-th/9608156}

\preref\SDG{J.F.\ Plebanski, 
J.\ Math.\ Phys.\ 16:2395 (1975)\semi
M.J. Duff, in {\it Proceedings of 1979 Supergravity Workshop},
ed. P. Van Nieuwenhuizen and D.Z.\ Freedman (North Holland, 1979)\semi
W.\ Siegel, Phys.\ Rev. D47:2504 (1993), hep-th/9207043}

\preref\Review{%
Z. Bern, L. Dixon and D.A. Kosower, to appear in
{\it Annual Reviews of Nuclear and Particle Science} (1996),
hep-ph/9602280}

\preref\DuffIsham{
M.J. Duff, C.J. Isham, Nucl.\ Phys.\ B162:271 (1980)}

\preref\SDYM{
L. Dolan, Phys.\ Rep.\ 109:1 (1984), and references therein}

\preref\Vafa{C. Vafa, preprint hep-th/9602022}

\preref\DouglasLi{M.R. Douglas and M. Li, preprint hep-th/9604041}

\preref\Kutasov{D. Kutasov and E. Martinec, preprint hep-th/9602049\semi
D. Kutasov, E. Martinec and M. O'Loughlin, preprint hep-th/9603116}


\noindent

$\null$
 
\vskip -.6 cm
  
\noindent hep-th/9611127 \hfill SLAC--PUB--7355

\hfill SWAT-96-142

\hfill UCLA/96/TEP/34

\hfill SACLAY-SPHT-T96/132

\vskip -2.0 cm 

\baselineskip 12 pt
\Title{\bf One-Loop Self-Dual and N=4 Super Yang-Mills}
 
\vskip -.2 cm 
  
\centerline{\ninerm Zvi Bern${}^{\sharp}$}
\baselineskip=13pt
\centerline{\nineit Department of Physics, UCLA, Los Angeles, CA
90095, USA}
\centerline{\tt bern@physics.ucla.edu}
\vglue 0.3cm
   
\centerline{\ninerm Lance Dixon${}^{\star}$}
\centerline{\nineit Stanford Linear Accelerator Center, 
Stanford University, Stanford, CA 94309, USA}
\centerline{\tt lance@slac.stanford.edu}
   \vglue 0.3cm
    
\centerline{\ninerm David C. Dunbar${}^{\dagger}$ }
\centerline{\nineit Department of Physics, University 
    of Wales Swansea, Swansea, SA2 8PP, UK }
\centerline{\tt d.c.dunbar@swan.ac.uk}
     
\vglue 0.2cm
\centerline{\ninerm and}
\vglue 0.2cm
\centerline{\ninerm David A. Kosower${}^{\ddagger}$}
\baselineskip12truept
\centerline{\nineit Service de Physique Th\'eorique,
Centre d'Etudes de Saclay}
\centerline{\nineit F-91191 Gif-sur-Yvette cedex, France}
\centerline{\tt kosower@spht.saclay.cea.fr}

\vskip 0.2truein
\baselineskip13truept
\centerline{\bf Abstract}

{\narrower We conjecture a simple relationship between the one-loop
maximally helicity violating gluon amplitudes of ordinary QCD
(all helicities identical) and those of $N=4$ supersymmetric Yang-Mills 
(all but two helicities identical).  Because the amplitudes in
self-dual Yang Mills have been shown to be the same as the maximally
helicity violating ones in QCD, this conjecture implies that they
are also related to the maximally helicity violating ones of $N=4$
supersymmetric Yang-Mills.  We have an explicit proof of the relation
up to the six-point amplitude; for amplitudes with more external legs,
it remains a conjecture.  A similar conjecture relates amplitudes
in self-dual gravity to maximally helicity violating $N=8$ supergravity
amplitudes.}

\vskip .7 cm 
\centerline{\it Submitted to Physics Letters B}

\vfil\vskip .2 cm
\noindent\hrule width 3.6in\hfil\break
${}^{\sharp}$Research supported in part by the US Department of
Energy
under grant DE-FG03-91ER40662 and in part by the
Alfred P. Sloan Foundation under grant BR-3222. \hfil\break
${}^{\star}$Research supported by the Department of
Energy under grant DE-AC03-76SF00515.\hfil\break
${}^{\dagger}$Research supported by PPARC,
the Leverhulme trust and EEC contract ERBCHRXCT920069.
\hfil\break
${}^{\ddagger}$Laboratory of the {\it Direction des Sciences de
la Mati\`ere\/}
of the {\it Commissariat \`a l'Energie Atomique\/} of
France.\hfil\break
       \eject

\baselineskip17pt

\noindent
{\bf 1. Introduction}

\noindent
The development of sophisticated techniques~[\use\Review]
for computing one-loop helicity amplitudes in four-dimensional 
gauge theories has allowed various workers to obtain
explicit expressions for a number of infinite sequences of such 
amplitudes~[\use\AllPlus,\use\Mahlon,\use\SusyFour,\use\SusyOne].
In particular, the nonvanishing maximally helicity
violating (MHV) one-loop gluon amplitudes in QCD (where all helicities
are identical) and in $N=4$ supersymmetric Yang-Mills (where all but two
helicities are identical) are remarkably simple.  This suggests
that they may possess an additional symmetry beyond the 
gauge symmetry.

At tree level, Nair [\use\Nair] has observed that the MHV $n$-gluon
amplitudes [\use\ParkeTaylor] (also known as Parke-Taylor amplitudes) may
be derived from a free-fermion Wess-Zumino-Witten model which contains an
infinite-dimensional symmetry algebra. (The construction was actually
for an $N=4$ supersymmetric gauge theory, but the superpartners do not
contribute at tree level, so the results also apply to ordinary QCD.)
Duff and Isham [\use\DuffIsham], and more recently Bardeen [\use\Bardeen], 
have pointed out that tree-level gluon currents with all 
identical helicities in ordinary QCD may be
obtained from self-dual Yang-Mills.  Selivanov has also produced
similar results using a different ansatz [\use\Selivanov].  
Self-dual Yang-Mills is the prototypical integrable model 
and as such possesses an infinite-dimensional symmetry 
algebra [\use\SDYM].
In a spacetime of signature $(2,2)$, it arises from the 
$N=2$ string~[\use\NTwoString].

Recently, Cangemi~[\use\Cangemi,\use\CangemiConf] and Chalmers and
Siegel~[\use\Siegel] showed that a connection between amplitudes in
self-dual Yang-Mills and the maximally helicity violating all-plus
helicity amplitudes in QCD continues to hold at one-loop.  Indeed, the
one-loop amplitudes generated by various self-dual Yang-Mills actions
[\use\DNS,\use\LCG,\use\Siegel] are identical to the QCD all-plus
helicity amplitudes.  It is intriguing that the action of Chalmers and
Siegel leads to a perturbatively solvable theory: the {\it only\/}
non-vanishing amplitudes in the perturbative expansion are the known
all-plus helicity one-loop amplitudes in QCD!  Bardeen has suggested
that an anomaly in the symmetry algebra determines the structure of
these amplitudes [\use\Bardeen].

In this paper we examine the relationship between 
the one-loop MHV amplitudes in $N=4$ supersymmetric Yang-Mills theory
and the all-plus helicity QCD amplitudes 
(i.e., the self-dual Yang-Mills amplitudes).  
We conjecture a `dimension shifting' relationship between the 
two sets of amplitudes, in which the all-plus amplitudes are given
essentially by evaluating the loop integration for the $N=4$ MHV 
amplitudes in a dimensions larger by four($D=8$). 
We have explicitly verified the conjecture for amplitudes with 
up to six external legs, and have evidence that it holds for 
an {\it arbitrary\/} number of external legs.  
A similar conjecture can be made to link the one-loop $n$-point amplitudes 
of self-dual gravity [\use\SDG,\use\Siegel] 
(the all-plus helicity graviton amplitudes), 
with MHV amplitudes in $N=8$ supergravity.  
We have verified this conjecture for the four-point amplitude. 
The underlying symmetry responsible for the simplicity 
of these amplitudes, and their relation to each other,
remains to be clarified.

\vskip 0.3 truecm
\noindent
{\bf 2. Preliminaries}

\noindent
We now review two basic tools necessary to present the conjecture,
namely color-ordering and the spinor helicity formalism.
Further details may be found in review articles
[\use\ManganoReview,\use\Review], whose normalizations and 
conventions we follow.

One-loop $SU(N_c)$ gauge theory amplitudes can be written in 
terms of independent color-ordered partial amplitudes multiplied 
by an associated color structure~[\use\TreeColor,\use\BKLoopColor].
As a simple example, the decomposition of
the four-gluon amplitude (with adjoint particles in the loop) is
$$
\eqalign{
{\cal A}_4 (\{ a_i, k_i, \pol_i\}) =
g^4 \sum_{\sigma} N_c \, & \Tr(T^{\as1}   T^{\as2} T^{\as3} T^{\as4} )
 A_{4;1}(\sig1,\sig2,\sig3,\sig4)
\cr
&+ g^4 \sum_{\rho}
\Tr(T^{a_{\rho(1)}} T^{a_{\rho(2)}} )
\Tr(T^{a_{\rho(3)}} T^{a_{\rho(4)}} ) 
A_{4;3}(\sig1,\sig2;\sig3,\sig4) \,, 
\cr}
\anoneqn
$$
where we have abbreviated the arguments of the `partial amplitudes',
$A_{n;j}$, by the labels $i$ of the legs and the $T^{a_i}$ are fundamental
representation matrices, normalized so that $\Tr(T^a T^b) =
\delta^{ab}$.  The $\rho$ and $\sigma$ permutation sums are over the
ones which alter the color trace structure.  The structure for any
number of legs is similar, with no more than two color traces appearing
in each term (at one loop).  
String theory suggests, and it has been proven in field
theory, that the $A_{n;j>1}$ may be obtained from $A_{n;1}$ by an
appropriate permutation sum [\use\BKLoopColor,\use\SusyFour,\use\Fermion].
Thus, we need only consider the $A_{n;1}$ --- they contain the 
information necessary to reconstruct the full one-loop amplitude, and 
any identity proven for the $A_{n;1}$ extends automatically to the
full amplitude.

The relations we find are for special choices of the external 
gluon helicities.
In the helicity formalism of Xu, Zhang and Chang [\use\SpinorHelicity] 
the gluon polarization vectors are expressed
in terms of Weyl spinors $\vert k^{\pm} \rangle$ as
$$\pol^{+}_\mu (k;q) =
{\sandmm{q}.{\gamma_\mu}.k
\over  \sqrt2 \spa{q}.k}\,, \hskip 2 cm  
\pol^{-}_\mu (k;q) =
{\sandpp{q}.{\gamma_\mu}.k
\over \sqrt{2} \spb{k}.q} \, ,
\anoneqn
$$
where $k$ is the gluon momentum and $q$ is an arbitrary null
`reference momentum' which drops out of final gauge-invariant
amplitudes.  The plus and minus labels on the polarization vectors
refer to the gluon helicities and we use the notation
$\langle ij \rangle\equiv  \langle k_i^{-} \vert k_j^{+} \rangle\, ,
[ij] \equiv \langle k_i^{+} \vert k_j^{-} \rangle$.
These spinor products are anti-symmetric and satisfy
$\spa{i}.j \spb{j}.i = 2 k_i \cdot k_j $.

When performing a calculation in dimensional regularization
[\use\DimReg] it is convenient to choose a scheme which is compatible
with the spinor helicity formalism.  We use the four-dimensional
helicity scheme [\use\Long] which is equivalent at one loop
to a helicity form of Siegel's dimensional reduction scheme
[\use\SusyReg].  The conversion to the standard $\overline{\rm MS}$ scheme
is discussed in refs.~[\use\Long,\use\SchemeConversion].

\vskip 0.3 truecm 
\noindent 
{\bf 3. Previously obtained amplitudes.} 
\vskip 0.1 truecm 

\noindent
The simplest one-loop QCD $n$-gluon helicity amplitude is the one with all 
identical helicities [\use\AllPlus,\use\Mahlon],
$$
\eqalign{
A_{n;1}^{\rm gluon}(1^+,2^+,\ldots,n^+)\ =\ -{i \over 48\pi^2}\,
\sum_{1\leq i_1 < i_2 < i_3 < i_4 \leq n}
{ {\rm tr}_-[i_1 i_2 i_3 i_4]
\over \spa1.2 \spa2.3 \cdots \spa{n}.1 } 
+ \ord(\eps) \,,}
\eqn\AllPlusOld
$$
where $\tr_-[i_1 i_2 i_3 i_4] \equiv {1\over 2}\tr[(1-\gamma_5)
\ksl_{i_1} \ksl_{i_2} \ksl_{i_3} \ksl_{i_4}]$ and the label
`gluon' denotes a gluon circulating in the loop. As indicated by the
`$+$' superscripts on the gluon labels, we have chosen the all-plus
helicity configuration; the all-minus helicity configuration is
related by parity. This amplitude contains no poles in the dimensional
regularization parameter $\eps=(4-D)/2$; it is both ultraviolet and
infrared finite.  
In a supersymmetric theory identical helicity amplitudes vanish by
a supersymmetry identity [\use\SWI]. This implies that the
contribution of a massless adjoint representation Weyl fermion or
complex scalar circulating in the loop is the same up to a statistics factor
[\use\TasiZvi,\use\AllPlus], 
$$
A_{n;1}^{\rm scalar}(1^+,2^+,\ldots,n^+) = 
- A_{n;1}^{\rm fermion}(1^+,2^+,\ldots,n^+) = 
A_{n;1}^{\rm gluon}(1^+,2^+,\ldots,n^+) \,,
\eqn\SusyIdentity
$$
where the labels `scalar' and `fermion' again refer to the
particle circulating in the loop.

The next simplest amplitude is the $N=4$ supersymmetric Yang-Mills
MHV amplitude [\use\SusyFour],
$$
\eqalign{
A_{n;1}^{ N=4}
&(1^+, 2^+, \ldots, i^-, \ldots, j^-, \ldots n^+) \cr
& \hskip -0.5 truecm
= {i \over (4\pi)^{2-\eps}}
{\spa{i}.j^4 \over \spa1.2 \spa2.3 \cdots \spa{n}.1}
\sum_{1\leq i_1<i_2\leq n} {1\over 4} \tr[\ksl_{i_1} \Psl_{i_1+1,i_2-1}
\ksl_{i_2} \Psl_{i_2+1,i_1-1}]\,
I^{D=4-2\eps}_{4:i_1,i_2} + \Ord(\eps) \,,\cr }
\eqn\SusyFourAmpOld
$$
where $P_{i,j} = \sum_{m=i}^{j} k_m$ and only legs $i$ and $j$ carry
negative helicity. (For notational convenience we take the labels
on the legs  mod $n$.) The box integral functions
$I^{D=4-2\eps}_{4:i_1,i_2}$ are depicted in
\fig\BoxFigure; $i_1,i_2$ on the integral function label the two
diagonally opposite massless legs.  The formal definition for the
$m$-point integral functions is
$$
I_m^{D}
\equiv i (-1)^{m+1} (4\pi)^{D/2} \int
{d^{D} \ell \over (2\pi)^{D} }
{1 \over \ell^2 (\ell - K_1)^2 \ldots 
(\ell - \sum_{i=1}^{m-1} K_i)^2 } \,,
\eqn\IntegralDef
$$
where the $K_i$ are the external momenta for the integral, which are
in general sums of adjacent external massless momenta $k_i$ 
for the amplitude, as indicated in fig.~\use\BoxFigure. 
The explicit forms of the box integral functions appearing in
eq.~(\use\SusyFourAmpOld), evaluated to $\ord(\eps^0)$ in terms 
of logarithms and dilogarithms, may be found in the appendices of
refs.~[\use\Integrals,\use\SusyFour].

\vskip -1.1 cm

\LoadFigure\BoxFigure{\baselineskip 13 pt
\noindent\narrower\ninerm  The scalar box integrals appearing in the
$N=4$ MHV amplitudes.}
{\epsfysize 1.in}
{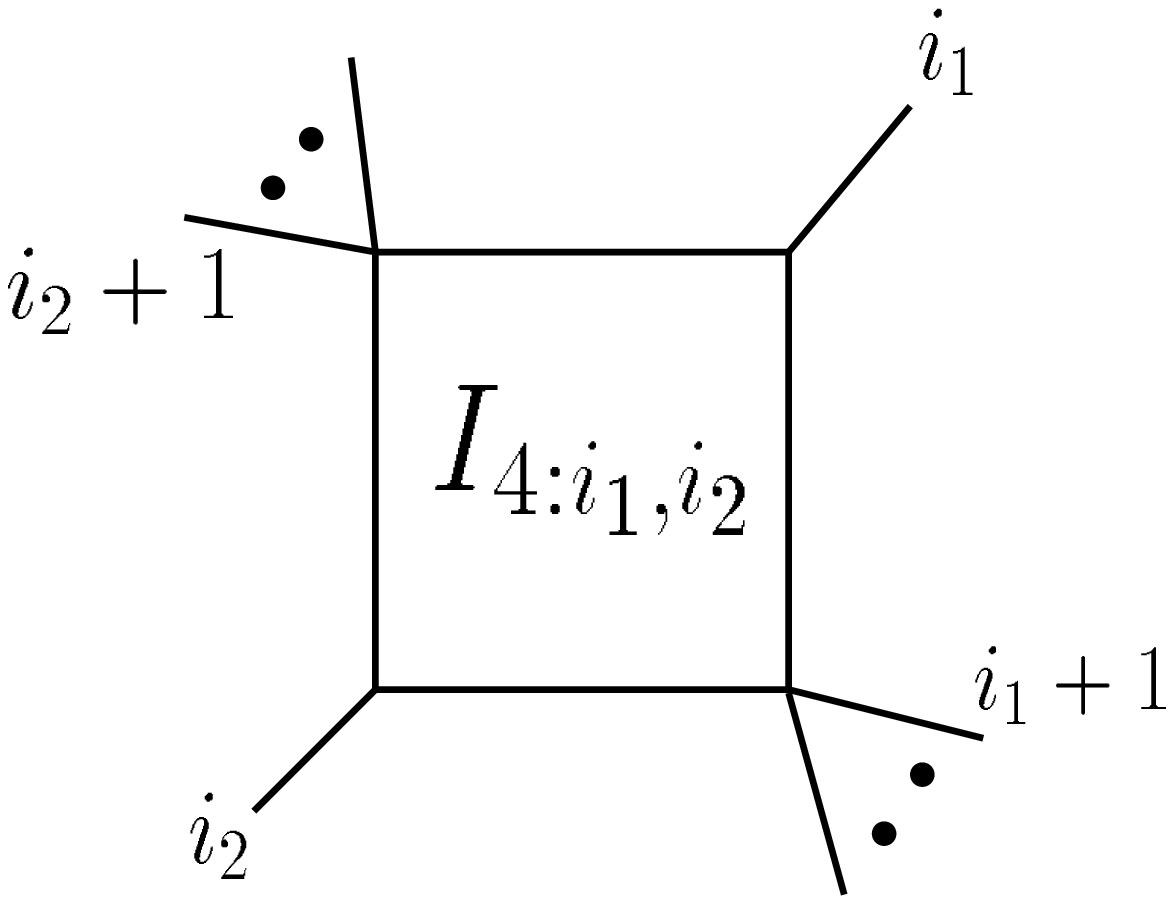}

The $N=4$ supersymmetric Yang-Mills MHV amplitudes (\use\SusyFourAmpOld)
have some features in common with the all-plus helicity QCD amplitudes
(\use\AllPlusOld).  Neither contains multi-particle poles.  The
appearance exclusively of two-particle poles is reminiscent of the
`Bethe ansatz' for integrable systems [\use\Bardeen].  On the other
hand, the $N=4$ supersymmetric Yang-Mills amplitudes contain infrared
singularities as well as logarithms and dilogarithms which are not
found in the all-plus helicity amplitudes.  In this paper we will
argue that up to an overall prefactor the two amplitudes are actually
the same after an appropriate shift of the dimension $D$ appearing
in the loop integrals~(\use\IntegralDef).

In refs.~[\use\Cangemi,\use\Siegel] it was shown that self-dual
Yang-Mills generates the same amplitudes as the all-plus helicity QCD
amplitudes.  These comparisons were done on the actions and Feynman
rules, so that the equivalence holds to all orders of the dimensional
regularization parameter, assuming that we are using a form of
dimensional regularization that modifies the dimension of the loop
momentum [\use\SusyReg,\use\Long], but preserves the number of
physical states to their $D=4$ values.  With this type of
regularization we can define a simple analytic continuation of self
dual-Yang-Mills (whose definition contains the four-dimensional
Levi-Civita tensor) 
in the dimensional regularization parameter.

\vskip 0.3 truecm 
\noindent 
{\bf 4. The conjecture.} 
\vskip 0.1 truecm 

The basic relationship we conjecture is,
$$
A^{\rm gluon}_{n;1} (1^+, 2^+, \ldots, n^+) = 
- 2 \eps (1-\eps) (4 \pi)^2 
\LB   {A^{N=4}_{n;1} (1^+, 2^+, \ldots, i^-, \ldots, j^-, \ldots,
  n^+) \over \spa{i}.j^4} 
    \biggr|_{D \rightarrow D+4} \RB\,,
\eqn\Conjecture
$$
where $D=4-2\eps$ and the dimension shift on the $N=4$ amplitude
takes $\eps \rightarrow \eps -2$ and $I_m^D \rightarrow
I_m^{D+4}$.  It leaves the external momenta and
helicities invariant (as well as the explicit prefactor of 
$\eps (1-\eps)$). 

One can motivate the conjecture in the $\e\to0$ limit
by recognizing that the box integral functions in the $N=4$
supersymmetric expression~(\use\SusyFourAmpOld) have a
common logarithmic ultraviolet divergence as $D\to8$, 
$I^{D=8-2\e}_{4:i_1,i_2} \sim 1/6\e$ as $\e\to0$,
which is canceled by the explicit $\e$ on the right-hand-side
of~(\use\Conjecture).  One then uses
$$
\eqalign{
\sum_{1\leq i_1<i_2\leq n} \tr[\ksl_{i_1} \Psl_{i_1+1,i_2-1}
\ksl_{i_2} \Psl_{i_2+1,i_1-1}]
\ &= \sum_{1\leq i_1<j<i_2<k\leq n} \tr[i_1 j i_2 k] \; 
    + \!\!\sum_{1\leq k<i_1<j<i_2\leq n} \tr[i_1 j i_2 k] \cr
 &= 2 \hskip -.2 cm  \sum_{1\leq i_1<i_2<i_3<i_4\leq n} \tr[i_1 i_2 i_3 i_4]
\cr}
\eqn\Motivation
$$
to see that that these terms generate the `even' 
terms in $A_{n;1}^{\rm scalar}$ (i.e., those terms obtained by 
neglecting the $\gamma_5$ in $\tr[(1-\gamma_5)\cdots]$ in
eq.~(\use\AllPlusOld)).  
On the other hand, one cannot check the `odd' ($\gamma_5$) terms
in this way; we shall see (for $n=5,6$) that they come from 
$\Ord(\eps)$ terms in~(\use\SusyFourAmpOld) 
which are promoted to $\Ord(\eps^0)$ through the dimension shift.
In other words, because it involves a shift in $\eps$, the dimension 
shift in (\use\Conjecture) only makes sense when the amplitudes 
are expressed to all orders in $\eps$.
The previously calculated amplitudes (\use\AllPlusOld) and
(\use\SusyFourAmpOld) are valid only through $\Ord(\eps^0)$, so we
must inspect the terms higher order in $\eps$ to fully check 
the conjecture.

The conjecture~(\use\Conjecture) may also be  
reformulated in terms of the loop momentum integration. 
The $D$-dimensional integration in
eq.~(\use\IntegralDef) may be broken up into four- and
$(-2\eps)$-dimensional parts, allowing us to define
$$
I_m^{D}[\mu^{2r}] 
\equiv i (-1)^{m+1} (4\pi)^{D/2} \int 
    {d^{4} p \over (2\pi)^{4} }
    {d^{-2\eps} \mu \over (2\pi)^{-2\eps} }
  { \mu^{2r} \over (p^2 - \mu^2) \ldots (
    (p-\sum_{i=1}^{m-1} K_i )^2 - \mu^2) } \,,
\eqn\MuIntegralDef
$$
where $\mu$ is the $(-2\eps)$-dimensional part of the original loop
momentum.  (We follow the standard prescription that the
$(-2\eps)$-dimensional subspace is orthogonal to the four-dimensional
one.)  Explicit evaluation of the $(-2\eps)$-dimensional parts of the
integrals relates the integrals with powers of $\mu^2$ in the
numerator to higher-dimensional integrals (see, for example, appendix
A.2 of ref.~[\use\Massive]), yielding
$$
I_m^{D=4-2\eps}
[\mu^{2r}] =  -\eps (1-\eps) \cdots (r-1-\eps) I_m^{D=4+2r-2\eps} \,.
\anoneqn
$$

With the definition of the integrals (\use\MuIntegralDef) we may reformulate
the conjecture (\use\Conjecture) as
$$
A^{\rm gluon}_{n;1} (1^+, 2^+, \ldots, n^+) = 
2 {A^{N=4}_{n;1} (1^+, 2^+, \ldots, i^-, \ldots, j^-, \ldots,  n^+)
[\mu^4]  \over \spa{i}.j^4} \,,
\eqn\ConjectureMu
$$
where the symbol `$[\mu^4]$' indicates that we insert an extra factor
of $\mu^4$ into every loop integrand before performing the integrals.

\vskip 0.3 truecm 
\noindent 
{\bf 5. Evidence for the conjecture.}

\noindent
We shall present evidence for the conjecture (\use\Conjecture), but
first let us address a seeming puzzle with it.  The all-plus helicity
amplitude is invariant under a cyclic relabeling of the legs, whereas
the cyclic invariance of the $N=4$ supersymmetric amplitude is not
obvious, because the two negative helicities break the manifest
invariance.  However, the cyclic symmetry of the $N=4$ MHV amplitude, up
to the overall prefactor of $\spa{i}.j^4$, follows from a
supersymmetry identity.  To prove this, use standard supersymmetry
identities [\use\SWI] to relate the $n$-gluon amplitude to the two
scalar, $(n-2)$ gluon amplitude.  After interchanging the two scalars,
which does not affect the amplitude, use the same supersymmetry
identities to obtain an amplitude with the negative helicity gluon in
a different position.  This argument works for the $N=4$ multiplet
because the two gluon helicity states are related by supersymmetry
(without using a CPT transformation).

We have verified the conjecture for the four-, five- and 
six-point amplitudes by explicitly calculating both sides
of eq.~(\use\Conjecture) to all orders in $\e$.
To calculate the all-plus helicity amplitudes we use the 
unitarity-based method recently reviewed in ref.~[\use\Review].  
In this method the amplitudes are constructed from cut loop 
momentum integrals, depicted in \fig\CutFigure,
$$
\eqalign{
A_{n;1}&(1, 2, \ldots, n)\Bigr|_{\rm cut}  =  \cr
& \hskip -3mm
\int\! {d^{4-2\eps}\ell\over (2\pi)^{4-2\eps}} \;
  {i\over \ell_1^2 } \,
 A_{m_2-m_1+3}^\tree (-\ell_1, m_1, \ldots, m_2,\ell_2) 
  \,{i\over \ell_2^2} \,
 A_{n-m_2+m_1+1}^\tree (-\ell_2 ,m_2+1,\ldots, m_1-1,\ell_1)
                          \biggr|_{\rm cut} \, ,\cr}
\eqn\TreeProductDef
$$
where $\ell$ is the loop momentum and $\ell_1$ and $\ell_2$ are the
momenta crossing the cut. 
This equation is valid only for the cut channel.  One then
reconstructs the complete amplitudes by finding a function which has
the correct cuts in {\it all\/} channels.

\vskip -.9 cm 
\LoadFigure\CutFigure{\baselineskip 13 pt
\noindent\narrower\ninerm  The cut amplitude corresponding to 
eq.~(\use\TreeProductDef).}
{\epsfysize 1.05in}
{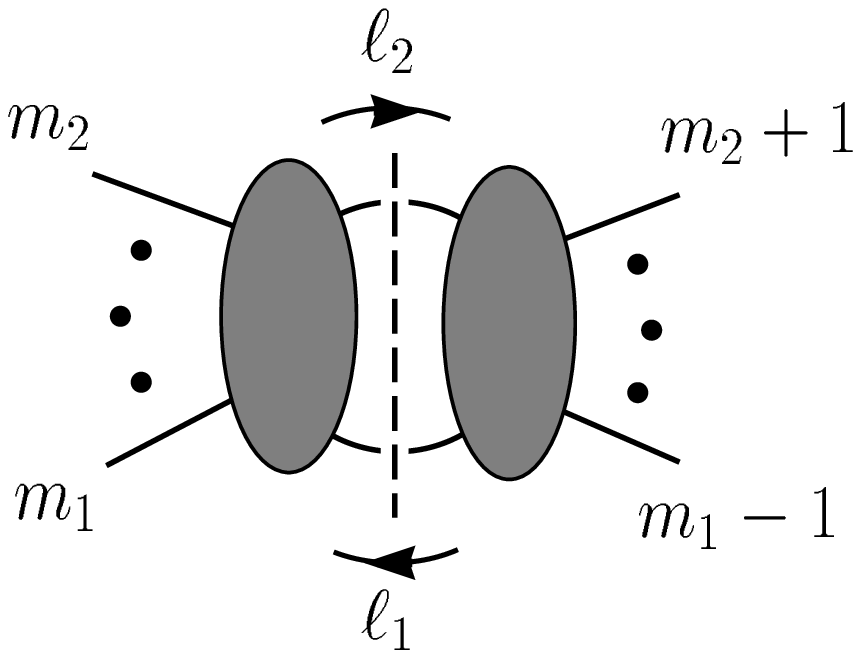}

{}From eq.~(\use\SusyIdentity), for the all-plus helicity amplitude we
only need calculate the case of a complex scalar circulating in loop.
Thus in eq.~(\use\TreeProductDef) we require 
the tree amplitudes with all-plus helicity gluons and two 
complex scalars, and the integration is over the momenta of
the complex scalars.  When working to all orders in $\e$,
we must use tree amplitudes that are valid for $D$-dimensional 
cut momenta.  In terms of the break-up of the loop momentum 
discussed in the previous section, the proper on-shell conditions 
on the cut legs are $\ell_1^2 - \mu^2 = 0$
and $\ell_2^2 - \mu^2 = 0$, where $\ell_1$ and $\ell_2$ are 
the four-dimensional components, and $\mu$ is the $(-2\eps)$-dimensional
component, of the loop momentum. 
For practical purposes we may think of $\mu^2$ as a mass that
gets integrated over.

Using recursive techniques [\use\Recursive,\use\Mahlon] we find 
$$
\eqalign{
A^{\rm tree}_4(-\ell_1^s,1^+,2^+,\ell_2^s) &=  i 
{\mu^2\spb1.2\over\spa1.2 [( \ell_1 -k_1)^2-\mu^2] }\,, \cr
A^{\rm tree}_5(-\ell_1^s,1^+,2^+,3^+,\ell_2^s ) &=
 i {\mu^2 \sum_{j=1}^{2} \spb{3}.j \sand{j^-}.{\lsl_1}.{1^-}\over
[(\ell_1-k_1)^2-\mu^2]  \spa1.2 \spa2.3 
[(\ell_2+k_3)^2-\mu^2] } \,, \cr
A^{\rm tree}_6(-\ell_1^s, 1^+, 2^+, 3^+, 4^+, \ell_2^s) & =  
i {1\over [(\ell_1 - k_1)^2 -\mu^2]
          \spa1.2 \spa2.3 \spa3.4 [(\ell_2 + k_4)^2 - \mu^2]}\cr
&\hskip 5mm\times \biggl[ \mu^2  
\sum_{j=1}^{3}\spb{4}.j\sandmm{j}.{\lsl_1}.{1}
- {\mu^4 \spb1.2\spa2.3\spb3.4\over (\ell_1 - k_1 - k_2)^2  
-\mu^2}\biggr]\,,
 \cr }
\anoneqn
$$
where the superscript $s$ on $\ell_1$ and $\ell_2$ indicates that
these are the scalar lines.

Integrating these tree amplitudes according to
eq.~(\use\TreeProductDef), and reconstructing the complete analytic 
form of the loop amplitudes we have
$$
\eqalign{
A_{4;1}^{\rm scalar}&(1^+,2^+,3^+,4^+) =
{ 2i \over  \spa1.2\spa2.3\spa3.4\spa4.1}
{\eps(1-\eps) \over (4\pi)^{2-\eps} }
\times s_{12}s_{23} I_4^{D=8-2\eps} \,,
\cr}
\eqn\Fourpt
$$
$$
\eqalign{
A_{5;1}^{\rm scalar}&(1^+,2^+,3^+,4^+,5^+) =
{ i \over  \spa1.2\spa2.3\spa3.4\spa4.5\spa5.1}
{\eps(1-\eps) \over (4\pi)^{2-\eps} }
\cr
&\times
\Bigl[
s_{23}s_{34} I_4^{(1),D=8-2\eps}
+s_{34}s_{45} I_4^{(2),D=8-2\eps}
+s_{45}s_{51} I_4^{(3),D=8-2\eps}
\cr
&+s_{51}s_{12} I_4^{(4),D=8-2\eps}
+s_{12}s_{23} I_4^{(5),D=8-2\eps}
+(4-2\eps) { \varepsilon (1,2,3,4) }I_5^{D=10-2\eps}
\Bigr] \,,
\cr}
\eqn\Fivept
$$
$$
\eqalign{
& A_{6;1}^{\rm scalar}(1^+, 2^+, 3^+, 4^+, 5^+, 6^+) =
 {i \over \spa1.2 \spa2.3 \spa3.4 \spa4.5 \spa 5.6 \spa6.1}
{\eps(1-\eps)\over(4\pi)^{2-\eps} }   \cr
& \hskip 2 cm \times
\hf \biggl[
- \hskip -.3 cm \sum_{1\leq i_1<i_2 \leq 6} \hskip -.2 cm
\tr[\ksl_{i_1} \Psl_{i_1+1, i_2-1}
\ksl_{i_2} \Psl_{i_2+1,i_1-1}] I_{4:i_1;i_2}^{D=8-2\eps}
+  (4-2\eps)\, \tr[123456] \, I_6^{D=10-2\eps} \cr
& \hskip 2 cm
+  (4-2\eps) \sum_{i=1}^6 \varepsilon(i+1, i+2, i+3, i+4)
I_{5}^{(i),D=10-2\eps} \biggr] \,, \cr  }
\eqn\Sixpt
$$
where $s_{ij} =  (k_i + k_j)^2$, the totally antisymmetric symbol
is defined by
$$
\varepsilon (i,j,m,n) \equiv 4i\varepsilon_{\mu\nu\rho\sigma}
        k_i^\mu k_j^\nu k_m^\rho k_n^\sigma
= \tr[\gamma_5 \ksl_i \ksl_j \ksl_m \ksl_n] \,, 
\anoneqn
$$
and $I_n^{(i)}$ denotes the scalar integral obtained by removing the
loop propagator between legs $i-1$ and $i$ from the $(n+1)$-point
scalar integral.
It is easy to verify that each of the amplitudes 
(\use\Fourpt)--(\use\Sixpt) properly reduces to the expression in
eq.~(\use\AllPlusOld), using values of the integrals in the 
$\eps \rightarrow 0$ limit, 
$$
\eps(1-\eps) I_4^{D=8-2\eps} \rightarrow {1\over 6}\,, \hskip 1.3 cm 
\eps(1-\eps) I_5^{D=10-2\eps} \rightarrow {1\over 24}\,, \hskip 1.3 cm 
\eps(1-\eps) I_6^{D=10-2\eps} \rightarrow  0 \,. 
\anoneqn
$$

We comment that these amplitudes may be converted to ones with a
massive loop simply by performing the shift $\mu^2 \rightarrow \mu^2
+m^2$ [\use\Massive].  Just as in the massless case, a supersymmetry
identity implies that the all-plus helicity amplitude depends only on
the number of statistics-weighted states circulating in the loop; thus
the above conversion~(\use\SusyIdentity) also applies for a massive
fermion in the loop.  One may convert these amplitudes from QCD to QED
simply by summing over permutations of the external legs.  

We now compare the all-plus helicity amplitudes
(\use\Fourpt)--(\use\Sixpt) with the $N=4$ MHV amplitudes.  The $N=4$
four-point amplitude was first calculated by Green, Schwarz and
Brink [\use\GSB] using the low energy limit of superstring theory.  We
obtained the five-point amplitude by slightly modifying the
string-based [\use\Long,\use\StringBased] calculation of
ref.~[\use\FiveGluon] to keep the terms higher order in $\eps$.  For
the six-point amplitudes we used a string-motivated diagrammatic
approach to ensure manifest supersymmetric cancellations
[\use\FiveGluon,\use\TasiZvi,\use\WeakInt], after which the diagrams
were evaluated numerically.  (Hexagon integrals $I_6$ with external
momenta restricted to four-dimensions are linear combinations of the
six pentagon integrals $I_5^{(i)}$~[\use\MVNV,\use\Integrals];
therefore we had to reduce the hexagon to pentagons before making any
comparison.)  In all these cases we find that the dimension-shifting
formula~(\use\Conjecture) is satisfied, thus proving the conjecture up
through $n=6$.

What evidence can we find for an arbitrary number of external legs?
We noted above that if we start with the $N=4$ supersymmetric
amplitudes (\use\SusyFourAmpOld) valid through $\Ord(\eps^0)$, perform
the dimension shift, and then take the $\eps\rightarrow 0$ limit, we
reproduce all `even' terms in the all-plus helicity 
amplitudes (\use\AllPlusOld).  This
check is not definitive since terms of $\Ord(\eps)$ can become terms
of $\Ord(\eps^0)$ under the dimension shift.  For example, present in
the five-point $N=4$ amplitude is the $\Ord(\eps)$ `odd' term
$$
-2\eps \,{ \varepsilon (1,2,3,4) }I_5^{D=6-2\eps} \,.
\anoneqn
$$
After shifting $D\rightarrow D+4$, and multiplying by the prefactor
$-\eps (1-\eps)$ this becomes
$$
-\eps(1-\eps) \times (4-2\eps) \varepsilon (1,2,3,4) I_5^{D=10-2\eps}\,,
\anoneqn
$$
which contributes at order $\eps^0$ because the integral is
ultraviolet divergent.  {}From the explicit forms of the
all-orders-in-$\eps$ five- and six-point amplitudes, it is clear that
the `odd' terms arise from integral functions not contributing through
order $\eps^0$ in $A^{N=4}$.

As a stronger check, we may appeal to the universal behavior
[\use\Factorization] of the amplitudes as kinematic invariants vanish.
Of particular utility is the behavior of amplitudes as two momenta
become collinear
[\use\ParkeTaylor,\use\ManganoReview,\use\AllPlus,\use\Review].  In
these limits an $n$-point amplitude must reduce to sums of
$(n-1)$-point amplitudes multiplied by `splitting functions' which are
singular in the collinear limit.  The constraints of factorization are
sufficiently powerful that in many cases one may obtain the correct
amplitude simply by finding a function that satisfies the constraints
[\use\AllPlus].  Since the conjecture (\use\Conjecture) holds for up
to six-point amplitudes, consistency of the collinear limits suggests
that it will continue to hold for higher-point
amplitudes.  This argument is not a proof either, given the possible
appearance of functions which are non-singular in all factorization limits; 
these limits do not constrain such functions.  An example of such
a function for the $n$-point amplitude (if $n$ is even) is
$$
{\tr[123\cdots n] \over \spa1.2 \spa2.3 \spa3.4 \cdots \spa{n}.1}
I_n^{D=n+4-2\eps} \,.
\anoneqn
$$
This function does appear in the six-point ($n=6$) amplitude 
(\use\Sixpt), but only at $\Ord(\eps)$.  While collinear factorization
does not prove the conjecture for $n>6$, it severely constrains terms
which violate it.

Another way to check the conjecture (\use\Conjecture) is to inspect
the cuts (to all orders in $\eps$) on both sides of the equation.
This is convenient since the cut of a one-loop amplitude is a product
of two tree amplitudes integrated over phase space.  Tree amplitudes
are in turn easier to manipulate than loop amplitudes.  The cut
relationship implied by the conjecture (\use\Conjecture) is shown
diagrammatically in \fig\CutProofFigure\ for the case where the two
negative helicities lie on the same side of the cut.  (It is
sufficient to check this case because of the supersymmetry identity
proving the cyclic symmetry of the $N=4$ MHV amplitudes.)  This may be
expressed as
$$
\eqalign{
  & \int\dlips
  \ A^\tree(-\ell_1^s,m_1^+,\ldots,m_2^+,\ell_2^s)
  \ A^\tree(-\ell_2^s,(m_2+1)^+,\ldots,(m_1-1)^+,\ell_1^s) 
=
\cr
  &  2 \int \dlips
\sum_{f} {\mu^4 \over \spa{i}.{j}^4 } 
A^\tree(-\ell_1^f, m_1^+, \ldots, i^-, \ldots, j^-, \ldots, m_2^+, \ell_2^f)
  \ A^\tree(-\ell_2^f, (m_2+1)^+, \ldots, (m_1-1)^+, \ell_1^f) \,,
\cr}
\eqn\Check
$$
where the summation is over the states $f$, 
of the $N=4$ multiplet and the integration, $\dlips$ is over
Lorentz invariant phase space with $\ell_1,\ell_2$  on-shell.

\vskip -.5 cm 
\LoadFigure\CutProofFigure{\baselineskip 13 pt
\noindent\narrower\ninerm Equality needed for conjecture to be true.
In the cut on the left, only scalars cross the cut; in the cut on
the right, the entire $N=4$ supersymmetry multiplet appears.}
{\epsfysize 1.2in}
{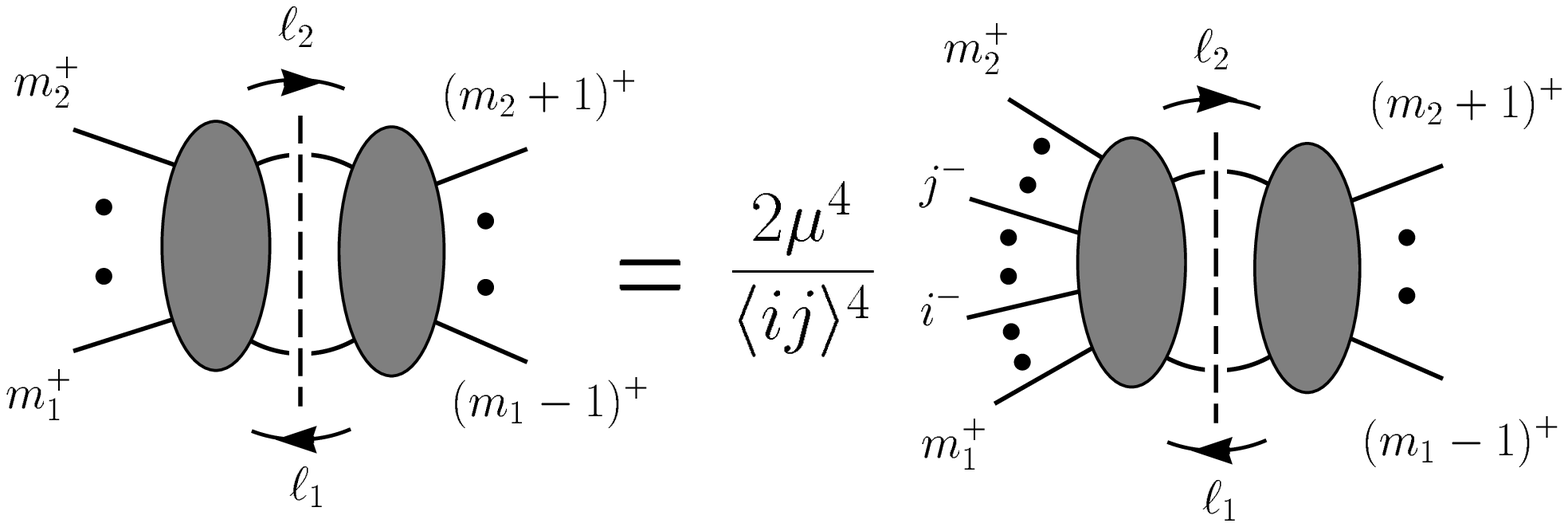}

A proof of this identity would lead directly to a proof of the
conjecture~(\use\Conjecture).  We offer no proof to all orders in
$\mu$; but as a first step, let us consider this equation to leading
order in $\mu^2$.  The leading order on both sides is $\mu^4$.  On the
$N=4$ side of the equation we can use the amplitudes to zeroth order
in $\mu^2$.  These cuts were evaluated (to obtain those terms in the
amplitudes which do not vanish as $\eps \rightarrow 0$) in ref. [\use\SusyFour]
with the result,
$$
\eqalign{
2\sum_{f}
& { \mu^4 \over \spa{i}.{j}^4 }
A^\tree(-\ell_1^f,m_1^+,\ldots, i^- \ldots j^-, \ldots, m_2^+,\ell_2^f)
  \ A^\tree(-\ell_2^f,(m_2+1)^+, \ldots, (m_1-1)^+, \ell_1^f ) \cr
& = 2{ \mu^4 \over \spa1.2 \spa2.3 \ldots \spa{(n-1)}.{n} \spa{n}.1 }
   { \spa{(m_1-1)}.{m_1} {\spa{\ell_1}.{\ell_2}}^2 \spa{m_2}.{(m_2+1)}
    \over \spa{(m_1-1)}.{\ell_1} \spa{\ell_1}.{m_1} \spa{m_2}.{\ell_2}
    \spa{\ell_2}.{(m_2+1)} } \cr
& =
-2{ \mu^4 \over \spa1.2 \spa2.3 \ldots \spa{(n-1)}.{n} \spa{n}.1 }\cr
& \hskip 2 cm \times
   { \tr_+[\lsl_1 \ksl_{m_1} \ksl_{m_1 - 1} \Psl_{m_1, m_2}\, \lsl_2 
      \ksl_{m_2 + 1} \ksl_{m_2} \Psl_{m_2 + 1, m_1 -1}]   
    \over [(\ell_1 - k_{m_1})^2 - \mu^2] [(\ell_1 + k_{m_1-1})^2 - \mu^2]
      [(\ell_2 + k_{m_2})^2 - \mu^2] [(\ell_2 - k_{m_2+1})^2 - \mu^2]} \,,\cr}
\eqn\SusyLeadingMu
$$
where we used $\ell_2 = \ell_1 - P_{m_1,m_2}
   = \ell_1 + P_{m_2+1,m_1-1}$ and $\ell_i^2 = 0$ on the cut.
We can now compare this result with the leading order in $\mu^2$ 
for the all-plus helicity case.  
Recursive techniques [\use\Recursive,\use\Mahlon] 
lead to the general form of the tree amplitudes for
$n$ plus-helicity gluons and two scalars,
$$
\eqalign{
A_n^{\rm tree}(-\ell_1^s,1^+,\ldots,n^+,\ell_2^s ) &=
i {\mu^2 \sum_{j=1}^{n-1} \spb{n}.{j}\sandmm{j}.{\ell_1}.1
       \over [(\ell_1-k_1)^2-\mu^2] \spa{1}.{2} \cdots\spa{(n-1)}.{n} 
    [(\ell_2+k_n)^2 -\mu^2]}  + \Ord(\mu^4) \,.
 \cr }
\anoneqn
$$
Using this expression to construct the cuts one reproduces
eq.~(\use\SusyLeadingMu), so that eq.~(\Check) is satisfied to leading
order in $\mu^2$.  (The overall factor of 2 arises because complex
scalars are composed of two states.) The agreement, even before
performing the phase-space integrals, suggests that, in general, on
the cuts the conjecture holds for the integrands.

\vskip 0.3 truecm 
\noindent 
{\bf 6. Gravity.} 
\vskip 0.1 truecm 

\noindent
String theory implies that gravity amplitudes are closely related to
gauge theory amplitudes.  This observation has been used to obtain
gravity amplitudes at both tree level [\use\BerendsGravity] and at
loop level [\use\Gravity] and suggests that one can find conjectures
similar to eq.~(\use\Conjecture), but for gravity.

Using the explicit results for four-graviton amplitudes obtained via
string-based calculations [\use\Gravity], extended to all-orders
in $\e$, we find
$$
\eqalign{
A^{\rm gravity}_4 (1^+, 2^+, 3^+, 4^+ ) 
&=
- 2 \eps (1-\eps)(2-\eps)(3-\eps)  (4 \pi)^4 \LB
{A^{N=8}_4 (1^-, 2^-,  3^+, 4^+ ) 
  \over \spa{1}.2^8} 
      \biggr|_{D \rightarrow D+8} \RB
\cr
&=
{2 A^{N=8}_4 (1^-, 2^-,  3^+, 4^+ )[\mu^8]
  \over \spa{1}.2^8} \,,
  \cr}
\eqn\GravityEquation
$$
where the amplitude on the left is the pure gravity all-plus helicity
amplitudes and the one on the right the $N=8$ supergravity amplitude.
As in the QCD case, the all-plus amplitude is independent of the massless
particle types circulating in the loop, but depends only on the number of
states in the loop.

Following the QCD case, we may conjecture that the relation in
eq.~(\use\GravityEquation) continues to hold for an arbitrary number
of external legs.  For gravity the one-loop amplitudes are not known
beyond four external legs.  One can, however, argue [\use\Siegel] that the
above amplitudes will also correspond to those for self-dual gravity
[\use\SDG].

\vskip 0.3 truecm 
\noindent 
{\bf 7. Speculations.}

\noindent
In this paper we have provided evidence that two infinite sequences of
maximally helicity violating gauge theory amplitudes, which at first
sight seem quite different, are in fact closely related to each other
through a ``dimension shift''.  Is this result just a
curiosity, or an indication of a deeper relation between a
non-supersymmetric theory (self-dual Yang-Mills) and a supersymmetric
one ($N=4$ super Yang-Mills)?  We cannot yet answer this question
directly.  It may prove profitable to pursue the connection mentioned
in the introduction, between maximal helicity violation and self-dual
Yang-Mills theory [\use\Bardeen,\use\Cangemi,\use\Siegel], since the
latter is known to possess an infinite-dimensional symmetry algebra
[\use\SDYM]. (See ref.~[\use\CangemiConf] for a review.)
In two-dimensional integrable models, which are related to self-dual
Yang-Mills theory through dimensional reduction, the extended symmetry
algebra is responsible for a lack of multi-particle poles in the
scattering amplitude.  Bardeen has emphasized that the absence of
multi-particle poles in the maximally helicity violating tree-level
currents is reminiscent of the Bethe ansatz [\use\Bardeen].

Thus it might be worthwhile to examine the other four-dimensional
gauge theory amplitudes that lack multi-particle poles.  The list of
such amplitudes is quite limited.  In non-supersymmetric QCD, beyond
one loop all amplitudes with six or more legs contain multi-particle
poles, as can be verified by checking their factorization onto a
product of two one-loop amplitudes.  On the other hand, the
nonvanishing maximally helicity violating amplitudes in supersymmetric
theories (those amplitudes with all but two helicities identical) do
not develop multi-particle poles, to all orders of perturbation
theory.  (The residues of the would-be poles vanish by supersymmetry
identities~[\use\SWI].)  The simplest of the one-loop MHV
supersymmetric amplitudes are the $N=4$ amplitudes, which is why we
chose to investigate their relationship to the self-dual Yang-Mills
amplitudes in this letter.  ($N=1$ partial amplitudes [\use\SusyOne]
are more complicated and certainly do not possess the cyclic
invariance of the $N=4$ amplitudes.)

Finally, we speculate whether to take seriously the appearance of
dimensions shifted upwards by four units (eight units for gravity) in
the relations we have found.  If we take $\eps\to0$ so that the
left-hand side of eq.~(\use\Conjecture) is in $D=4$, we find that the
self-dual gauge amplitudes are related to the one-loop ultraviolet
divergences of an $N=4$ supersymmetric gauge theory in $D=8$.  ($N=4$
refers to the number of $D=4$ supersymmetries.)  Coincidentally, such
theories have recently been considered in the context of certain
$(7+1)$-brane configurations in string theory (also known as
compactifications of $F$ theory on $K3$) where they describe the
low-energy world-volume theory [\use\Vafa,\use\DouglasLi].  The
corresponding theory for the gravity relation~(\use\GravityEquation)
would be $N=8$ supergravity in $D=12$, which happens to be the
``critical dimension'' for $F$ theory [\use\Vafa].  Perhaps it is also
relevant that self-dual theories in four-dimensions (with signature
$(2,2)$) have been proposed for the world-volume dynamics of $F$
theory [\use\Kutasov].  At this stage, though, it is safest to say
that the underlying reason for the relationships (\use\Conjecture) and
(\use\GravityEquation) remains to be clarified.

\noindent
{\bf Acknowledgements.}

We thank D. Cangemi and M. Douglas for a number of useful discussions.
Z.~B. and L.~D. thank the Aspen Center for Physics where part of this
work was performed.  Z.~B. and D.~D. also thank the Centre d'Etudes de
Saclay, and L.~D. thanks Rutgers University, for support and
hospitality.

\baselineskip13.8pt

\listrefs

\end